\def\Qem{{$Q_{\rm em}$}}

 \def\Z{{\bf Z}}

\def\EE{E$_8\times$E$_8^\prime$}
\def\Eo{E$_8$}

\def\Uan{U(1)$_{\rm an}$}

\def\PG{$P_\Gamma$}

\def\Sp{${\cal S}$}
\def\Hvec{${\cal V}$}
\def\bffive{{\bf 5}}
\def\bffiveb{$\overline{\bf 5}$}
\def\bffiveh{${\bf 5}^H$}
\def\bffivebh{$\overline{\bf 5}^H$}
\def\bften{{\bf 10}}
\def\bftenb{$\overline{\bf 10}$}
\def\bftenh{${\bf 10}^H$}
\def\bftenbh{$\overline{\bf 10}^H$}

\documentclass[aps,nofootinbib,showpacs,showkeys,preprint
]
{revtex4}

\usepackage{epsf,epsfig,subfigure,axodraw,graphicx,amsmath,amssymb}
\usepackage{color}

\begin{document}
 \begin{flushright}
{\tt SNUTP 06-014\\
KIAS-P06069}
\end{flushright}
\title{\Large\bf Harmless R-parity violation from
$\Z_{12-I}$ compactification of \EE\ heterotic string
 }
\author{
Ian-Woo Kim$^{(a)}$\footnote{iwkim@phya.snu.ac.kr}, Jihn E.
Kim$^{(a)}$\footnote{jekim@phyp.snu.ac.kr} and Bumseok
Kyae$^{(b)}$\footnote{bkyae@kias.re.kr} }
\address{
 $^{(a)}$Department of Physics and Astronomy and Center for Theoretical
 Physics, Seoul National University, Seoul 151-747, Korea\\
 $^{(b)}$School of Physics, Korea Institute for Advanced Study,
207-43 Cheongryangri-dong, Dongdaemun-gu, Seoul 130-722, Korea}


\begin{abstract}
In a recent $\Z_{12-I}$ orbifold model, an approximate $Z_2$
symmetry which forbids the baryon number violating operators up to
sufficiently high orders is found. The dimension-4 $\Delta B\ne 0$
operators of the MSSM fields occur at dimension 10. The effective
dimension-5 $\Delta B\ne 0$ operators derived from these are
harmless if some VEVs of neutral singlets are O($\frac{1}{10}$)
times the string scale. The main reason for forbidding these
$\Delta B\ne 0$ operators up to such a high order is the large
order $N=12$ of $\Z_N$ since the $H$-momentum rule is $(-1,1,1)$
mod (12, 3, 12). For a lower order $N<12$, the $\Delta B\ne 0$
operators would appear at lower dimensions.
\end{abstract}

 \maketitle

\def\bfst{{\bf 16}}

\section{introduction}

The main reason for imposing the R-parity in the minimal
supersymmetric standard model (MSSM) is to forbid the dangerous
$\Delta B\ne 0$ superpotential terms. As a bonus, the exact
R-parity ensures an absolutely stable lightest supersymmetric
particle (LSP) as a candidate for cold dark matter. The R-parity
is a simple discrete symmetry in the MSSM to forbid dangerous
(renormalizable) $\Delta B\ne 0$ operators. However, as for the
condition on proton longevity, other discrete symmetries in
addition to the R-parity can be possible. All possible candidates
are classified in Refs.
 \cite{ibross, Dreiner:2005rd}. In this paper, we search for a scheme to
obtain such a discrete symmetry in compactifications of \EE\
heterotic string.

The well known R-parity in SO(10) grand unified theory (GUT) is by
assigning --1 for the spinor \bfst\ and +1 for the vector {\bf 10}.
This kind of spinor-vector disparity can be adopted in the {\it
untwisted sector} of heterotic string also, in particular in the
phenomenologically attractive \EE\ heterotic string \cite{GroHar}.
Let us consider only the \Eo\ part for an illustration. The {\it
untwisted sector}  massless matter spectrum in \Eo\ can be $P^2=2$
weights distinguished by the spinor or the vector property
\begin{align}
 {\cal S}: ([++++++++]) \quad
 {\cal V}: (\underline{\pm1~\pm1~0~0~0~0~0~0})\nonumber
\end{align}
where $\pm$ represents $\pm\frac12$, the notation [ ] means
including even number of sign flips inside the bracket, and the
underline means permutations of the entries on the underline. Since
two spinors in the group space can transform as a tensor in the
group space, cubic Yukawa couplings arising from the untwisted
sector involving two spinors are of the form \Sp \Sp \Hvec, which
can be used to assign a kind of matter parity. For this scheme to
work, 48 fermions of the three families (including the singlet
neutrinos to generate neutrino masses) must belong to the untwisted
sector  \Sp\ and Higgs doublets must belong to the untwisted sector
\Hvec. This requirement is nontrivial as one finds that the
standard-like models of \cite{standlike} do not satisfy this
condition. On the other hand, a part of the spectrum of the
$\Z_{12}$ model of \cite{Kim:2006hv} satisfies this condition. But
this spinor-vector disparity condition alone is not enough to
guarantee a harmless discrete symmetry in supersymmetric standard
models because there are twisted matter, and Yukawa couplings are
more constrained than this simple spinor--vector disparity.

The twisted matter is more complicated because of the form
(internal momenta) plus (shift vectors), $P+kV$ ($k=1,2,\cdots ,
N-1$ in a $\Z_{N}$ orbifold). Except in the $\Z_2$ orbifold, there
appear fractional numbers such as $\frac13,\frac14,\cdots$. So, in
general it is very much involved to find a discrete parity if one
includes interactions of twisted matter.

GUTs allow harmless proton decay via gauge interactions if the GUT
scale $M_{\rm GUT}$ is greater than $10^{15}$ GeV. String models in
addition contain proton decay operators through Yukawa couplings
which are measured by the string scale $M_S$ being considered to be
O(10) larger than the GUT scale. If these proton decay operators
occur at dimension 4 and 5, they can dominate over the proton decay
operators via gauge interactions \cite{SakYan}. The R-parity forbids
dimension-4 $\Delta B\ne 0$ operators, but does not forbid
dimension-5 $\Delta B\ne 0$ operators. So, in some supergravity GUT
models  dimension-5 $\Delta B\ne 0$ operators are considered to be
the dominant ones of proton decay, predicting $p\to K+ (\rm
antilepton)$.\footnote{One of simple ways to forbid the dimension-4
and -5 $\Delta B\ne 0$ operators in supergravity is to introduce a
$U(1)$ R-symmetry. However, it should be broken to a discrete
symmetry in orbifold string compactifications.  It is known that in
a supergravity model \cite{Dreiner:2005rd}, a $Z_6$ symmetry also
forbids such $\Delta B\ne 0$ operators. In Ref.
\cite{anomalousU(1)}, for instance, an anomalous \Uan\ is employed
for the R-parity and also for the proton longevity.} In string
compactifications, however, the dimension-5 operators are considered
to be dangerous because the coefficients are considered to be O(1)
in general \cite{ibross}. So, if we introduce a kind of matter
parity, it must work very ingeniously to forbid the dimension-4 and
dimension-5 $\Delta B\ne 0$ operators.

If a parity is introduced, it is better to be a discrete gauge
symmetry \cite{discreteg}, otherwise large gravitational corrections
such as through wormhole processes may violate it. Even if the
discrete symmetry is broken as we consider in this paper, it is
better to be a discrete gauge symmetry to free us from gravitational
corrections. The reason for anticipating a broken parity in string
compactifications is that we will embed the parity in a global U(1)
which is not an exact symmetry in string compactifications. But the
breaking of the parity will be considered to be {\it harmless} if
the $\Delta B\ne 0$ operators derived from breaking that parity is
sufficiently suppressed or suppressed by masses greater than
$10^{17}$ GeV since anyway $\Delta B\ne 0$ operators are present in
the gauge sector of string GUTs.

Suppose an approximate global symmetry U(1)$_\Gamma$ and its
discrete $Z_2$ subgroup \PG. It is a kind generalizing the
R-parity. In this paper, we use the word $\lq$R-parity' even
though we restrict the discussion to the matter parity.

The parity we consider cannot be put in a general form but must be
discussed based on specific models. Restricting to specific models
is obvious because the approximate global symmetry U(1)$_\Gamma$
must be given in a specific model. This leads us to the discussion
of the specific model, Ref. \cite{Kim:2006hv}.

\section{Continuous U(1) symmetries}

We observe that a discrete subgroup of the anomalous \Uan\ of Ref.
\cite{Kim:2006hv} is not good for the R-parity because neutral
singlets carry even and odd \Uan\ charges.  A good candidate for
housing a part of R-parity is U(1)$_X$ of flipped SU(5). In Table
\ref{tb:flipobs}, we list $X$ charges of the non-exotic fields as
subscripts. We do not list exotic fields and $X=0$ singlet fields.
\begin{table}
\begin{center}
\begin{tabular}{|c|c|c||c|c|c|}
\hline  Visible states & $SU(5)\times U(1)_X$& $\Gamma$& Visible
states & $SU(5)\times U(1)_X$& $\Gamma$
\\
\hline
 $(\underline{+----};+++)$ & ${\bf {5}}_{3}^{L}(U_3)$ &3 &
$(\underline{1,0,0,0,0};\frac{-1}{3},0^2)$ &
$3\cdot{\bf{5}}_{-2}^L(T4^0)$
 &--2
\\
$(\underline{+++--};+--)$ & $\overline{\bf 10}_{-1}^{L}(U_3)$ &--1 &
$(\underline{-1,0,0,0,0};\frac{-1}{3},0^2)$ &  $2\cdot\overline{\bf
 5}_{2}^L(T4^0)$ &2
\\
$({+++++};+++)$ & ${\bf 1}_{-5}^{L}(U_3)$ & --5 &
$(\underline{+----};+~0~0)$ & $2\cdot {\bf 5}_{3}^L(T6)$&4
\\
$(\underline{-1,0,0,0,0};-1,0,0)$ & $\overline{\bf 5}_{ 2}^{L}(U_2)$
& 2 & $(\underline{+++--};+~0~0)$ &$4\cdot\overline{{\bf
10}}_{-1}^L(T6)$ &--2, --1
\\
$(\underline{+----};+--)$ & ${\bf {5}}_{3}^{L}(U_1)$ & 3&
$({+++++};+~0~0)$ &$2\cdot{\bf 1}_{-5}^L(T6)$ & --6
\\[0.2em]
$(\underline{+++--};+++)$ & $\overline{\bf 10}_{-1}^{L}(U_1)$ &--1 &
  $(\underline{++++-};-~0~0)$ &
$2\cdot\overline{{\bf 5}}_{-3}^L(T6)$ &--4
\\[0.2em]
$(+++++;+--)$ & ${\bf 1}_{-5}^{L}(U_1)$ & --5&
  $(\underline{++---};-~0~0)$ & $3\cdot{\bf 10}_{1}^L(T6)$ &2
\\
$(\underline{+----};\frac{-1}{6}~0~0)$ & ${\bf{5}}_{3}^L(T2^0)$ &3 &
 $({-----};-~0~0)$ & $2\cdot {\bf{1}}_{5}^L(T6)$ &6
\\
$({+++++};\frac{-1}{6}~0~0)$ & ${\bf 1}_{-5}^L(T2^0)$ &--5 & & &
\\
\hline
\end{tabular}
\end{center}
\caption{Relevant visible sector chiral fields. The multiplicity is
shown as the coefficients of representations. + and -- represent
$+\frac12$ and $-\frac12$, respectively. The $\Gamma$ values of --2
and --1 for $\overline{\bf 10}_{-1}^{L}(T6)$ are those for the
vectorlike ones and for the $t$-quark family,
respectively.}\label{tb:flipobs}
\end{table}

The key representations of the flipped SU(5), i.e. $SU(5)\times
U(1)_X$, are
\begin{align}
&{\rm matter}: \overline{\bf 10}_{-1},\
{\bf 5}_{3},\ {\bf 1}_{-5} \label{eq:matter}\\
&{\rm Higgs}:\left\{\begin{array}{l} \overline{\bf 5}_{2},\ {\bf
5}_{-2},\quad{\rm electroweak\ scale}\\
 \overline{\bf 10}_{-1}^H,\ {\bf 10}_{1}^H,
 \quad{\rm GUT\ scale}
 \end{array}\right.\label{eq:Higgs}
\end{align}
If we restrict to matter and the electroweak scale Higgs fields
only, the $Z_2$ subgroup of U(1)$_X$ is a good candidate for the
prime source of the R-parity since matter fields, Eq.
(\ref{eq:matter}), carry odd $X$ quantum numbers and the
electroweak scale Higgs \bffive\ and \bffiveb\ (the first line of
Eq. (\ref{eq:Higgs})) carry even $X$ quantum numbers. Singlets
with $X=0\ (\pm 5)$ are neutral (\Qem=$\pm 1$) singlets. To break
the flipped SU(5) down to the standard model, the GUT scale Higgs
\bftenh\ and \bftenbh\ develop GUT scale VEVs.

In fact, the $Z_2$ subgroup of U(1)$_X$ distinguishes the spinor
or the vector origin of our spectrum since we assign
\begin{equation}
X=(-2,-2,-2,-2,-2,~0,~0,~0).
\end{equation}
For example, \bftenb\ from spinor of the form
$(\underline{--+++}\cdots)$ has an odd $X$, while \bffive\ from
vector of the form $(\underline{1~0~0~0~0}\cdots)$ has an even $X$.
This shows that both \bftenh\ and \bftenbh\ having odd $X$s.
Therefore, for the light fields we have a perfect definition for the
R-parity, but including heavy fields \bftenh\ and \bftenbh\ make us
rethink on the R-parity.

The basic difference between Higgs \bftenbh\ and matter \bftenb\
in $T6$ is that the former forms a vectorlike representation with
\bftenh\ in $T6$ and is removed at the GUT scale, while the latter
remains as a chiral one. From Table \ref{tb:flipobs}, there appear
four \bftenb$_{-1}^L$ s and three \bften$_{1}^L$ s in the twisted
sector $T6$. One unmatched \bftenb$_{-1}^L$ is interpreted as
belonging to the $t$-quark family. The gauge sector of four
\bftenb$_{-1}^L$s and three \bften$_{1}^L$s has the symmetry
$U(4)\times U(3)$. We factor out one U(1) belonging to the
$t$-quark family. The remaining symmetry of three \bftenbh\ and
three \bftenh\ is $U(1)_{V_{10}}\times SU(3)_L\times SU(3)_R\times
U(1)_A$. The $U(1)_A$ symmetry is broken by the anomaly and hence
we do not consider it. We are interested only in U(1)s because we
will assign the parity as a subgroup of a U(1). Thus, we do not
consider the nonabelian symmetry $SU(3)_L\times SU(3)_R$. Then, we
are left with the {\it global} symmetry $U(1)_{V_{10}}$. This
choice of anomaly free $U(1)_{V_{10}}$ is consistent with the
discrete gauge symmetry \cite{discreteg}. We define $V_{10}$
charges of \bftenh\ and \bftenbh\ are +1 and --1, respectively.
Matter \bftenb\ corresponding to the $t$-quark family carries the
vanishing $V_{10}$. Of course, this global symmetry
$U(1)_{V_{10}}$, not protected in string models, is broken by
Yukawa couplings.

From $T6$ sector in Table \ref{tb:flipobs}, we also find
\bffive$_3^L$ s and \bffiveb$_{-3}^L$ s carrying odd-$X$ quantum
numbers. Again, we call these two vectorlike pairs as \bffiveh\
and \bffivebh, for which we define another vector global symmetry
$U(1)_{V_{5}}$. Let $V_{5}$ charges of the vectorlike \bffiveh\
and \bffivebh\ be +1 and --1, respectively. On the other hand,
matter \bffive s which are chiral, carry the vanishing $V_{5}$
charge. Again, this global symmetry $U(1)_{V_{5}}$ is broken by
Yukawa couplings in our string compactification. For two
vectorlike pairs of \bffive\ and \bffiveb\
 of $T4$, carrying even-$X$ quantum numbers, we can consider such a
 U(1) symmetry but the $X$ charges are already even and we do not
 need another manipulation for these even-$X$ vectorlike pairs.

Before continuing discussion on $V_{10}$ and $V_5$ charges, let us
briefly comment on vectorlike representations of exotics. There are
two kinds of exotics. One kind is E-exotics which carry
$X=\pm\frac52$ charges, and hence they are \Qem$=\pm\frac12$
exotics. The other kind is G-exotics which carry SU(5) charges also:
${\bf 5}_{+\frac12}$ and $ \overline{\bf 5}_{-\frac12}$. Being
vectorlike, the E-exotics and G-exotics can be assigned respective
$U(1)_V$ quantum numbers as discussed in the preceding paragraph.
But they are more restricted than the global charges we discussed
for $U(1)_{V_{10}}$ and $U(1)_{V_{5}}$. The $U(1)_V$ symmetry for
exotics is identical to U(1)$_{\rm em}$ and it is not broken. The
electromagnetic charges of G-exotics are  $\pm\frac16$ for colored
ones in $ \overline{\bf 5}_{-\frac12}$ and ${\bf 5}_{+\frac12}$ and
$\pm\frac12$ for the doublet members in $ \overline{\bf
5}_{-\frac12}$  and ${\bf 5}_{+\frac12}$. The color singlet bound
states composed of colored ones of G-exotics carry
\Qem$=\pm\frac12$. Therefore, all color singlet exotics, elementary
and composite, carry \Qem$=\pm\frac12$. Because of the exact
U(1)$_{\rm em}$, the lightest \Qem$=\pm\frac12$ exotics are
absolutely stable. Therefore, we separate out the exotics from the
rest integer charged sector and do not consider the vectorlike
exotics anymore.

Going back to the vectorlike representations of \bftenh, \bftenbh,
\bffiveh, and \bffivebh, let us define a global charge $\Gamma$ as
\begin{equation}
\Gamma=X+V_{10}+V_{5}.\label{U1gamma}
\end{equation}
These $\Gamma$ charges are shown in Table \ref{tb:flipobs}. The
U(1)$_X$ symmetry is a gauge symmetry and exact, and
U(1)$_{V_{10}}$ and U(1)$_{V_{5}}$ are global symmetries and
approximate. When we break these gauge and global symmetries
spontaneously by one direction in the U(1) spaces, the gauge
symmetry is considered to be broken and a global symmetry remains.
The surviving global symmetry is U(1)$_\Gamma$. This is the
so-called 't Hooft mechanism \cite{tHooft}. In our case, the gauge
symmetry U(1)$_X$ is broken by the VEVs \bftenh\ and \bftenbh.
Because \bftenh\ carries $\Gamma=2$ and \bftenbh\ carries
$\Gamma=-2$, VEVs of \bftenh\ and \bftenbh\ do not break the $Z_2$
subgroup of U(1)$_\Gamma$. At this level, the continuous symmetry
is broken
\begin{equation}
U(1)_\Gamma\underset{\langle{\bf 10}^H \rangle}\longrightarrow
{\rm P}_\Gamma .\label{Pgamma}
\end{equation}

\section{Harmless R-parity as a discrete subgroup of U(1)$_\Gamma$}

Of course, the continuous global symmetry U(1)$_\Gamma$ is not
exact, being broken by superpotential terms. But, $\lq\lq$Is the
discrete subgroup \PG\ of U(1)$_\Gamma$ respected by all
superpotential terms?" It is not so. However, this is {\it
harmless} in the proton decay problem. To show this, let us note
possible superpotential terms in the MSSM, generating  $\Delta
B\ne 0$ operators,
\begin{align}
&D=4:\quad u^cd^cd^c, \label{D3uud}\\
&D=5:\quad qqql,\quad u^cu^cd^ce^+\label{D3qqql}
\end{align}
where $q$ and $l$ are quark and lepton doublets, respectively. The
dimension-4 operator of Eq. (\ref{D3uud}) alone  does not lead to
proton decay, but that term together with the $\Delta L\ne 0$
superpotential $qd^cl$ leads to a very fast proton decay and the
product of their couplings must satisfy a very stringent constraint,
$<10^{-26}$. The $D=5$ operators in (\ref{D3qqql}) are not that much
dangerous, but still the couplings must satisfy constraints,
$<10^{-8}$ \cite{SakYan,ibross}.

Therefore, we must forbid the $D=4$ operator. So let us look for
possibilities of generating the $u^cd^cd^c$ superpotential term.
 In the flipped SU(5), it is contained in
\bffive$_{3}$\bftenb$_{-1}$\bftenb$_{-1}$, which is however
forbidden by the SU(5)$\times$U(1)$_X$ gauge symmetry. In the
flipped SU(5), the $D=4,5$ operators are generated by
\begin{align}
D=4:&&\nonumber\\
&\quad [d^cd^cu^c]_F,\
 [qd^cl]_F\Leftarrow \langle\overline{\bf 10}^H\rangle\
\overline{\bf 10}\ \overline{\bf 10}\ {\bf 5},
\end{align}\label{baryon}
and
\begin{align}
&D=5:\nonumber\\
&\ \  \begin{array}{ll} O_1=[qqql]_F\Leftarrow \overline{\bf 10}\
\overline{\bf 10}\ \overline{\bf 10}\ {\bf 5},
 &O_2=[u^cu^cd^ce^+]_F
\Leftarrow{\bf 5}\ {\bf 5}\ \overline{\bf 10}\ {\bf 1}\\
O_3=[qqqH_d]_F\Leftarrow \langle{\bf 10}^H\rangle\ \overline{\bf
10}\ \overline{\bf 10}\ \overline{\bf 10}\ \overline{\bf 5}_2,
 & O_4=[qu^ce^+H_d]_F
\Leftarrow\langle{\bf 10}^H\rangle\ \overline{\bf
10}\ {\bf 5}\ {\bf 1}\ \overline{\bf 5}_{2}\\
O_5=[llH_uH_u]_F\Leftarrow \langle\overline{\bf 10}^H\rangle
\langle \overline{\bf 10}^H\rangle\ {\bf 5}\ {\bf 5}\
 {\bf 5}_{-2}\ {\bf 5}_{-2},
 &  O_6=[lH_dH_uH_u]_F \Leftarrow
 \langle \overline{\bf 10}^H\rangle\ {\bf
5} \  \overline{\bf 5}_2\ {\bf 5}_{-2}\ {\bf 5}_{-2}
\end{array}\label{Opd5}
\end{align}

In our model, even with including the possibilities of multiplying
a number of neutral singlets, we find that some of the above
operators are forbidden up to very high orders.

Since there are numerous neutral singlets which can acquire GUT or
string scale VEVs, we must consider all higher order terms also.
$u^c$ appears from ${\bf 5}_{3}$ in $U_3, U_1$ and $T2$. $d^c$
appears from $\overline{\bf 10}_{-1}$ in $U_3, U_1$ and $T6$. Note
that matter $\overline{\bf 10}_{-1}$ in $T6$ is the chiral one and
does not carry a $V_{10}$ charge, and it is the $\Gamma=-1$ part
out of four \bftenb s in $T6$. At $D=3$ level, there does not
appear $u^cd^cd^c$ due to the flipped SU(5) gauge symmetry before
considering any superstring property. One may consider $u^cd^cd^c
\cdot$(neutral singlets). Neutral singlets carry $X=0$ and hence
the possibility $u^cd^cd^c \cdot$(neutral singlets) is forbidden.
On the other hand, ${\bf 5}_3\overline{\bf 10}_{-1}\overline{\bf
10}_{-1} \overline{\bf 10}_{-1}$ is allowed by the gauge symmetry.
Giving a GUT scale VEV to one of $\overline{\bf 10}_{-1}$s, we
obtain the term $u^cd^cd^c$. It is a very dangerous R-parity
violating term. The $\overline{\bf 10}_{-1}$ obtaining a GUT scale
VEV is \bftenbh. So, effectively, we need a coupling
\bffive$\cdot$\bftenb$\cdot$\bftenb$\cdot$\bftenbh\ which breaks
the R-parity \PG\ (\bffive\ and \bftenb\ carry   $P_\Gamma=1$
while \bftenbh\ carries  $P_\Gamma=0$ according to (\ref{U1gamma})
and (\ref{Pgamma})). Thus the flipped SU(5) model at the
supergravity level is in jeopardy if the couplings of the form
${\bf 5}_3\overline{\bf 10}_{-1}\overline{\bf 10}_{-1}
\overline{\bf 10}_{-1}\cdot$(neutral singlets) are allowed. The
only cure of this problem at the {\it supergravity level} is
imposing the R-parity {\it by assumption}. But {\it superstring
models are free from such an assumption except for choosing the
vacuum.} It must be shown that such dangerous terms are
effectively forbidden in a superstring model, since the prime
motivation toward supersymmetry and superstring was to understand
the hierarchy of order $10^{-26}$.

We checked the coupling ${\bf 5}_3\overline{\bf
10}_{-1}\overline{\bf 10}_{-1} \overline{\bf 10}_{-1}$ attached with
singlets, and we do not find any such term up to $D=9$ with the
application of the program of Ref. \cite{ChKimKim}. This invites for
a careful check of the individual $H$-momenta of the fields so that
the computing time is drastically reduced. The needed $H$-momenta
are
\begin{align}
&U_1: (-1,0,0),\quad U_2: (0,1,0),\quad U_3:
(0,0,1),\nonumber\\
&\textstyle T2:(\frac{-1}{6},\frac46,\frac16),\quad
 T4:(\frac{-1}{3},\frac13,\frac13),\quad
T6:(\frac{-1}{2},0,\frac12)
\end{align}
Since ${\bf 5}_3$ has three possible locations $U_1,U_3$ and $T2$,
and $\overline{\bf 10}_{-1}$ has three possible locations
$U_1,U_3$ and $T6$, the possible combinations to be considered are
18 if we require one $\overline{\bf 10}_{-1}$ must be in $T6$.
These are tabulated in Table \ref{tb:dim5}.
\begin{table}[t]
\begin{center}
\begin{tabular}{|cccc|c|c||cccc|c|c|}
\hline  ${\bf 5}_3$  & $\overline{\bf 10}_{-1}$& $\overline{\bf
10}_{-1}$& $\overline{\bf 10}_{-1}$ & $H$-mom.& $N$ &
 ${\bf 5}_3$  & $\overline{\bf 10}_{-1}$& $\overline{\bf 10}_{-1}$
& $\overline{\bf 10}_{-1}$ & $H$-mom. & $N$
\\
\hline
 $U_1$ & $U_1$ &$U_1$ &
$T6$ & $(\frac{-7}{2},0,\frac12)$ &  11
 &$U_3$  & $U_3$ & $U_3$ & $T6$&$(\frac{-1}{2},0,\frac72)$
&  11
\\
 $U_1$ & $U_1$ &$U_3$ & $T6$ & $(\frac{-5}{2},0,\frac32)$
&  11
 &$U_3$  & $U_3$ & $T6$ & $T6$&$(-1,0,3)$ &
 12
\\
$U_1$ & $U_1$ & $T6$ & $T6$ & $(-3,0,1)$ &  12 & $U_3$  & $T6$ &$T6$
&$T6$ &$(\frac{-3}{2},0,\frac52)$ &  11
\\
$U_1$ & $U_3$ & $U_3$ & $T6$ & $(\frac{-3}{2},0,\frac52)$ &  11
&$T2$ &$U_1$ & $U_1$& $T6$&$(\frac{-8}{3},\frac23,\frac23)$ &
 10
\\
$U_1$ & $U_3$ & $T6$ & $T6$ &$(-2,0,2)$ & 12 &$T2$ &$U_1$ &$U_3$
&$T6$ &$(\frac{-5}{3},\frac23,\frac53)$ &  10
\\
$U_1$ & $T6$ &$T6$& $T6$ &$(\frac{-5}{2},0,\frac32)$ & 11 &$T2$
&$U_1$ & $T6$&$T6$ & $(\frac{-13}{6},\frac23,\frac76)$ &
 11
\\
$U_3$ & $U_1$ &$U_1$ & $T6$ &
 $(-\frac{5}{2},0,\frac{3}{2})$
&  11  &$T2$& $U_3$& $U_3$&$T6$
&$(\frac{-2}{3},\frac23,\frac{8}{3})$ &  10
\\
$U_3$ & $U_1$ & $U_3$&
  $T6$ & $(\frac{-3}{2},0,\frac52)$
&  11  &$T2$& $U_3$& $T6$&$T6$
&$(\frac{-7}{6},\frac23,\frac{13}{6})$ &  11
\\
$U_3$ & $U_1$ &$T6$ & $T6$ & $(-2,0,2)$ &  12 &$T2$& $T6$& $T6$&
$T6$
  &$(\frac{-5}{3},\frac23,\frac53)$
&  10
\\
\hline
\end{tabular}
\end{center}
\caption{\label{tb:dim5} Possible assignments of ${\bf 5}_3$ and
$\overline{\bf 10}_{-1}$s.  $N$ denotes the dimension of the lowest
dimensional $\Gamma$ breaking operators.}
\end{table}
Now a function of neutral singlets must be found so that the total
$H$-momentum together with those of Table \ref{tb:dim5} becomes
(--1, 1, 1) mod (12, 3, 12).

The neutral singlets appear from the sectors of $U_2, T2, T4,$ and
$T6$.  The numbers of neutral singlets from $U_2,T2,T4,$ and $T6$
are nonnegative $u_2,t=\sum t_i, f=\sum f_i,$ and $s=\sum s_i$,
respectively. The subscripts in the twisted sector distinguishes
the neutral singlets due to the oscillator contributions. All
neutral singlets in $T6$ carry oscillators and there are four
kinds of oscillators in neutral singlets in $T6$
\cite{Kim:2006hv}; thus we consider $s_i\ (i=1,\cdots,4)$. In
$T2$, neutral singlets appear with or without oscillators. There
are seven different $H$-momenta due to their oscillator
contributions; thus $t_i\ (i=1,\cdots,7)$. In $T4$, neutral
singlets have four different oscillator contributions; thus $f_i\
(i=1,\cdots,4)$. Thus, in $T2, T4$, and $ T6$ sectors, in addition
we must add oscillator contributions for neutral singlets
\begin{align}
T2:&\
 (0,0,0)t_1,\ (1,0,0)t_2,\  (0,1,0)t_3,\  (0,0,-1)t_4,\nonumber\\
 &\  (2,0,0)t_5,\    (0,0,-2)t_6,\  (1,0,-1)t_7\\
T4:&\   (0,0,0)f_1,\  (1,0,0)f_2,\  (0,-1,0)f_3,\  (0,0,-1)f_4,\\
T6:&\textstyle\   (1,0,0)s_1,\ (-1,0,0)s_2,\ (0,0,1)s_3,\
(0,0,-1)s_4.
\end{align}
Thus, we require three equations from the entries of $H$-momenta
\begin{align}
(0,&1,0)u_2\textstyle+(-\frac16,\frac23,\frac16) t+(-\frac13,
\frac13,\frac13) f+(-\frac12,0,\frac12) s\nonumber\\
&+
 (0,0,0)t_1+(1,0,0)t_2+ (0,1,0)t_3+ (0,0,-1)t_4\nonumber\\
 &+  (2,0,0)t_5+  (0,0,-2)t_6+ (1,0,-1)t_7\nonumber\\
&+(0,0,0)f_1+ (1,0,0)f_2+ (0,-1,0)f_3+ (0,0,-1)f_4\nonumber\\
&\textstyle\   +(1,0,0)s_1+(-1,0,0)s_2+
(0,0,1)s_3+(0,0,-1)s_4\nonumber\\
 &+(\rm the\ entries\ of\ H\
momenta)=(-1,1,1)\quad mod.\ (12,3,12).
\end{align}
The first entry equation is
\begin{align}
\textstyle -\frac16( t_1&-5t_2+t_3+t_4-11t_5+t_6-5t_7)
\textstyle -\frac13( f_1-2f_2+ f_3+ f_4 )\nonumber\\
&\textstyle-\frac12(-s_1+3s_2+s_3+s_4) +({\rm the\ first\ entry\ of\
H\ momenta})=-1,\ {\rm mod.}\ 12\label{Eqfirst}
\end{align}
The third entry equation is
\begin{align}
\textstyle \frac16( t_1&+t_2+t_3-5t_4+t_5-11t_6-5t_7)
\textstyle +\frac13( f_1+f_2+ f_3-2 f_4 )\nonumber\\
&\textstyle+\frac12(s_1+s_2+3s_3-s_4)+({\rm the\ third\ entry\ of\
H\ momenta})=+1,\ {\rm mod.}\ 12.\label{Eqthird}
\end{align}
The second entry equation is
\begin{align}
 u_2+&\textstyle\frac23( t_1+t_2+5t_3+t_4+t_5+t_6+t_7)
\textstyle +\frac13( f_1+f_2-2 f_3+ f_4 )\nonumber\\
& +({\rm the\ second\ entry\ of\ H\ momenta})=+1,\ {\rm mod.}\
3.\label{Eqsecond}
\end{align}
Adding Eqs. (\ref{Eqfirst}) and (\ref{Eqthird}), we obtain
\begin{align}
 t_2&-t_4+2t_5-2t_6+f_2-f_4
+s_1-s_2+s_3-s_4 \nonumber\\
& +({\rm sum\ of\ the\ 1st\ and\ 3rd\ entries\ of\ H\ momenta})=0,\
{\rm mod.}\ 12.\label{Sumfrth}
\end{align}

Consider the first case ($U_1U_1U_1T6$) of Table \ref{tb:dim5}. We
find a minimum order solution as $u_2=2,  t_2=3$, which would give
a dimension-9 operator. By the computer program, we checked that
there is no operators up to dimension-9. So, this solution must be
forbidden by the gauge symmetry. To check this, let us note that
the gauge U(1) charges of these singlets are \cite{Kim:2006hv}
\begin{align}
&u_2 : (0^8)(1,1;0^6)',\nonumber\\
 &\textstyle t_2:({\bf
1}_0;-\frac23,\frac12,\frac12) (\frac13,\frac13;0^6)'\nonumber
\end{align}
The gauge charge $(0^8)(2,2;0^6)'$ of two $u_2$s, i.e. two $s^u$s of
\cite{Kim:2006hv}, cannot be canceled by three $t_2$s. Thus, even if
the $H$-momentum rule allows it, the gauge invariance forbids it.
Another $H$-momentum solution $u_2=2, t_1=1, t_2=1, t_5=1$ does not
satisfy the gauge invariance either.

We looked for gauge invariant solutions satisfying Eqs.
(\ref{Eqfirst},\ref{Eqsecond},\ref{Eqthird}). The restriction from
the $H$-momentum rule saved computing time and we find that the
lowest order $\Delta B\ne 0$ operators occur at $D=10$. There are
thirty-four operators at $D=10$.
\begin{align}
W=&\ T2_5 U1_{\overline{10}} U1_{\overline{10}} T6_{\overline{10}}
\left\{
\begin{array}{c}
C^0_7 C^0_5 C^0_6 C^0_1 s^0_6 s^0_8
+C^-_3 C^0_5 C^0_6 C^+_1 s^0_6 s^0_8 \\
+C^0_7 C^0_5 C^0_6 C^0_1 s^+_3 s^-_2
+C^-_3 C^0_5 C^0_6 C^+_1 s^+_3 s^-_2
\end{array}
\right\} \nonumber \\
& + T2_5 U1_{\overline{10}} U3_{\overline{10}} T6_{\overline{10}}
\left\{
\begin{array}{c}
C^0_7 C^0_6 C^0_6 C^0_1 s^0_5 s^0_7
+C^-_3 C^0_6 C^0_6 C^+_1 s^0_5 s^0_7 \\
+C^0_7 C^0_6 C^0_6 C^0_1 s^+_2 s^-_3
+C^-_3 C^0_6 C^0_6 C^+_1 s^+_2 s^-_3 \\
+C^0_7 C^0_5 C^0_5 C^0_2 s^0_6 s^0_8
+C^-_3 C^0_5 C^0_5 C^+_2 s^0_6 s^0_8 \\
+C^0_7 C^0_5 C^0_5 C^0_2 s^+_3 s^-_2
+C^-_3 C^0_5 C^0_5 C^+_2 s^+_3 s^-_2 \\
+C^0_7 C^0_5 C^0_6 C^0_4 s^0_5 s^0_8
+C^-_3 C^0_5 C^0_6 D^+_2 d^+_2 s^-_2 \\
+C^-_3 C^0_5 C^0_6 D^+_2 d^+_3 s^-_3
+C^0_7 C^0_5 C^0_6 C^0_6 s^0_5 s^0_3 \\
+C^0_7 C^0_5 C^0_5 C^0_6 s^0_8 s^0_1
\end{array}
\right\} \nonumber \\
& +T2_5 U3_{\overline{10}} U3_{\overline{10}} T6_{\overline{10}}
\left\{
\begin{array}{c}
C^0_7 C^0_5 C^0_6 C^0_2 s^0_5 s^0_7
+C^-_3 C^0_5 C^0_6 C^+_2 s^0_5 s^0_7 \\
+C^0_7 C^0_5 C^0_6 C^0_2 s^+_2 s^-_3
+C^-_3 C^0_5 C^0_6 C^+_2 s^+_2 s^-_3
\end{array}
\right\} \nonumber \\
& +T2_5 T6_{\overline{10}} T6_{\overline{10}} T6_{\overline{10}}
\left\{
\begin{array}{c}
C^0_7 C^0_6 C^0_6 C^0_1 s^0_5 s^0_7
+C^-_3 C^0_6 C^0_6 C^+_1 s^0_5 s^0_7 \\
+C^0_7 C^0_6 C^0_6 C^0_1 s^+_2 s^-_3
+C^-_3 C^0_6 C^0_6 C^+_1 s^+_2 s^-_3 \\
+C^0_7 C^0_5 C^0_5 C^0_2 s^0_6 s^0_8
+C^-_3 C^0_5 C^0_5 C^+_2 s^0_6 s^0_8 \\
+C^0_7 C^0_5 C^0_5 C^0_2 s^+_3 s^-_2
+C^-_3 C^0_5 C^0_5 C^+_2 s^+_3 s^-_2 \\
+C^0_7 C^0_5 C^0_6 C^0_4 s^0_5 s^0_8
+C^-_3 C^0_5 C^0_6 D^+_2 d^+_2 s^-_2 \\
+C^-_3 C^0_5 C^0_6 D^+_2 d^+_3 s^-_3
+C^0_7 C^0_5 C^0_6 C^0_6 s^0_5 s^0_3 \\
+C^0_7 C^0_5 C^0_5 C^0_6 s^0_8 s^0_1
\end{array}
\right\} \label{d=4}
\end{align}

Thus, for  Table \ref{tb:dim5} the couplings ${\bf 5}_3\overline{\bf
10}_{-1}\overline{\bf 10}_{-1} \overline{\bf 10}_{-1}$ are of the
form
\begin{equation}
W\sim \left(\frac{\langle S^0\rangle}{M_S}\right)^{6}
\frac{\langle\overline{\bf 10}^H\rangle}{M_S} u^cd^cd^c
\end{equation}
where $M_S$ is the string scale close to O(10) times the GUT
scale. The upper bound of the coefficient is the squareroot of
$10^{-26}$ since $10^{-26}$ is on the product of coefficients of
two effective operators  $[d^cd^cu^c]_F$ and $[qd^cl]_F$ out of
the same coupling $\langle\overline{\bf 10}^H\rangle\
\overline{\bf 10}\ \overline{\bf 10}\ {\bf 5}$ in flipped SU(5).
Thus a coefficient of $u^cd^cd^c$ less than $10^{-13}$ is easily
achievable for $\langle S^0\rangle\sim(\frac{1}{100})M_S$ with
$\langle \overline{\bf 10}^H \rangle  \sim M_{\rm GUT}$. The
number $\frac{1}{100}$ is understood as an average number if we
choose the overall coefficient as 1. However, it should be noted
that some singlet VEVs can be much smaller than $10^{-2}$ and the
overall coefficient can be a relatively small number,\footnote{If
the nonrenormalizable couplings appear as connected cubic
diagrams, dimension-10 couplings would involve an extra factor of
(cubic couplings)$^7$ power. For cubic couplings of O($\frac15$),
the average VEV ratio can be made small to O($\frac{1}{10}$).} in
which case other singlet VEVs can be closer to $M_S$. This shows
that the dangerous $D=4$ operators of Eq. (10) are not harmful in
some vacua, $\langle S^0\rangle\sim(\frac{1}{100})M_S$, of the
$\Z_{12-I}$ compactification of the heterotic string
\cite{Kim:2006hv}. This proof also shows that the operators $O_1$
of Eq. (\ref{Opd5}) is perfectly harmless since one requires the
coefficient of $O_1$ being less than $10^{-8}$. The operator $O_3$
is not dangerous since the $\Delta B\ne 0$ process with $O_3$ also
needs the second operator of Eq. (8).

Thus, for dangerous $\Delta B\ne 0$ processes we are left with $O_2$
of Eq. (\ref{Opd5}). The phenomenological bound on the coefficient
of dimension-5 operator $O_2$ is taken as order $10^{-8}$. The
relevant flipped SU(5) term \bffive~\bffive~\bftenb~${\bf 1}$ needs
$U_1,U_3,T2$ (for matter \bffive), $U_1,U_3,T6$ (for matter
\bftenb), and $U_1,U_3,T2$ (for matter {\bf 1}). Possible cases are
54, which are listed in Table \ref{tb:O2}.
\begin{table}[t]
\begin{center}
\begin{tabular}{|cccc|c|c||cccc|c|c||cccc|c|c|}
\hline  ${\bf 5}_3$  & ${\bf 5}_3$& $\overline{\bf 10}_{-1}$& ${\bf
1}_{-5}$ & $H$-mom.& $N$&  ${\bf 5}_3$  & ${\bf 5}_3$&
$\overline{\bf 10}_{-1}$& ${\bf 1}_{-5}$ & $H$-mom.& $N$&  ${\bf
5}_3$ & ${\bf 5}_3$& $\overline{\bf 10}_{-1}$& ${\bf 1}_{-5}$ &
$H$-mom. & $N$
\\
\hline  $U_1$ & $U_1$ &$U_1$ & $U_1$ & $(-4,0,0)$ & 12
 &$U_1$  & $T2$ & $U_1$ & $U_1$&$(\frac{-19}{6},
 \frac{2}{3},\frac{1}{6})$ & 11 &$U_3$
  & $T2$ & $U_1$ & $U_1$&$(\frac{-13}{6},
 \frac{2}{3},\frac{7}{6})$ & 11
\\
$U_1$ & $U_1$ &$U_1$ & $U_3$ & $(-3,0,1)$       & 12  &
$U_1$ & $T2$  &$U_1$ & $U_3$ & $(\frac{-13}{6},\frac{2}{3},\frac{7}{6})$&11 &
$U_3$  & $T2$ & $U_1$ & $U_3$& $(\frac{-7}{6},\frac{2}{3},\frac{13}{6})$&11
\\
$U_1$ & $U_1$& $U_1$ & $T2$ &$(\frac{-19}{6},\frac{2}{3},\frac{1}{6})$ &11 &
$U_1$ & $T2$ & $U_1$ & $T2$ &$(\frac{-7}{3},\frac{4}{3},\frac{1}{3})$  &10 &
$U_3$ & $T2$ & $U_1$ & $T2$ &$(\frac{-4}{3},\frac{4}{3},\frac{4}{3})$  &10
\\
$U_1$ & $U_1$ & $U_3$& $U_1$&$(-3,0,1)$                                &12 &
$U_1$ & $T2$  & $U_3$& $U_1$&$(\frac{-13}{6},\frac{2}{3},\frac{7}{6})$ &11 &
$U_3$ & $T2$  & $U_3$& $U_1$&$(\frac{-7}{6},\frac{2}{3},\frac{13}{6})$ &11
\\
$U_1$ & $U_1$ & $U_3$& $U_3$&$(-2,0,2)$                                &12 &
$U_1$ & $T2$  & $U_3$& $U_3$&$(\frac{-7}{6},\frac{2}{3},\frac{13}{6})$ &11 &
$U_3$ & $T2$  & $U_3$& $U_3$&$(\frac{-1}{6},\frac{2}{3},\frac{19}{6})$ &11
\\
$U_1$ & $U_1$ & $U_3$& $T2$ &$(\frac{-13}{6},\frac{2}{3},\frac{7}{6})$ &11 &
$U_1$ & $T2$  & $U_3$& $T2$ &$(\frac{-4}{3},\frac{4}{3},\frac{4}{3})$  &10 &
$U_3$ & $T2$  & $U_3$& $T2$ &$(\frac{-1}{3},\frac{4}{3},\frac{7}{3})$  &10
\\
$U_1$ & $U_1$ & $T6$ & $U_1$&$(\frac{-7}{2},0,\frac12)$                &11 &
$U_1$ & $T2$  & $T6$ & $U_1$&$(\frac{-8}{3},\frac{2}{3},\frac{2}{3})$  &10 &
$U_3$ & $T2$  & $T6$ & $U_1$&$(\frac{-5}{3},\frac{2}{3},\frac{5}{3})$  &10
\\
$U_1$ & $U_1$ & $T6$ & $U_3$&$(\frac{-5}{2},0,\frac32)$                &11 &
$U_1$ & $T2$  & $T6$ &$U_3$ &$(\frac{-5}{3},\frac{2}{3},\frac{5}{3})$  &10 &
$U_3$  & $T2$ & $T6$ & $U_3$&$(\frac{-2}{3},\frac{2}{3},\frac{8}{3})$  &10
\\
$U_1$ & $U_1$ & $T6$ & $T2$ &$(\frac{-8}{3},\frac{2}{3},\frac{2}{3})$  &10 &
$U_1$ & $T2$  & $T6$ & $T2$ &$(\frac{-11}{6},\frac{4}{3},\frac{5}{6})$ & 9 &
$U_3$ & $T2$  & $T6$ & $T2$ &$(\frac{-5}{6},\frac{4}{3},\frac{11}{6})$ & 9
\\
$U_1$ & $U_3$ &$U_1$ & $U_1$&$(-3,0,1)$                                &12 &
$U_3$ & $U_3$ &$U_1$ & $U_1$&$(-2,0,2)$                                &12 &
$T2$  & $T2$  &$U_1$ & $U_1$&$(\frac{-7}{3},\frac{4}{3},\frac{1}{3})$  &10
\\
$U_1$ & $U_3$ &$U_1$ & $U_3$&$(-2,0,2)$                                &12 &
$U_3$ & $U_3$ &$U_1$ & $U_3$&$(-1,0,3)$                                &12 &
$T2$  & $T2$  &$U_1$ & $U_3$&$(\frac{-4}{3},\frac{4}{3},\frac{4}{3})$  &10
\\
$U_1$ & $U_3$ & $U_1$ & $T2$&$(\frac{-13}{6},\frac{2}{3},\frac{7}{6})$ &11 &
$U_3$ & $U_3$ & $U_1$ & $T2$&$(\frac{-7}{6},\frac{2}{3},\frac{13}{6})$ &11 &
$T2$  & $T2$ & $U_1$ & $T2$&$(\frac{-3}{2},2,\frac{1}{2})$             &9
\\
$U_1$ & $U_3$ & $U_3$ & $U_1$ &$(-2,0,2)$                              &12 &
$U_3$ & $U_3$ & $U_3$ & $U_1$ &$(-1,0,3)$                              &12 &
$T2$  & $T2$  & $U_3$ & $U_1$&$(\frac{-4}{3},\frac{4}{3},\frac{4}{3})$& 10
\\
$U_1$ & $U_3$ &$U_3$& $U_3$ &$(-1,0,3)$                                &12 &
$U_3$ & $U_3$ &$U_3$& $U_3$ &$(0,0,4)$                                 &12 &
$T2$  & $T2$  &$U_3$& $U_3$ &$(\frac{-1}{3},\frac{4}{3},\frac{7}{3})$  &10
\\
$U_1$ & $U_3$ &$U_3$ & $T2$ & $(\frac{-7}{6},\frac{2}{3},\frac{13}{6})$&11 &
$U_3$ & $U_3$ &$U_3$ & $T2$ & $(\frac{-1}{6},\frac{2}{3},\frac{19}{6})$&11 &
$T2$  & $T2$ & $U_3$ & $T2$ &$(\frac{-1}{2},2,\frac{3}{2})$            &9
\\
$U_1$ & $U_3$ & $T6$ & $U_1$ &$(\frac{-5}{2},0,\frac32)$               &11 &
$U_3$ & $U_3$ & $T6$ & $U_1$ &$(\frac{-3}{2},0,\frac{5}{2})$           &11 &
$T2$  & $T2$  & $T6$ & $U_1$ &$(\frac{-11}{6},\frac{4}{3},\frac{5}{6})$& 9
\\
$U_1$ & $U_3$ & $T6$ & $U_3$ &$(\frac{-3}{2},0,\frac52)$               &11 &
$U_3$ & $U_3$ & $T6$ & $U_3$ &$(\frac{-1}{2},0,\frac72)$               &11 &
$T2$  & $T2$  & $T6$ & $U_3$ &$(\frac{-5}{6},\frac{4}{3},\frac{11}{6})$& 9
\\
$U_1$ & $U_3$ & $T6$ &  $T2$ & $(\frac{-5}{3},\frac{2}{3},\frac{5}{3})$&10 &
$U_3$ & $U_3$ & $T6$ &  $T2$ & $(\frac{-2}{3},\frac{2}{3},\frac{8}{3})$ &10 &
$T2$  & $T2$ & $T6$ & $T2$   &$(-1,2, 1)$                               &8
\\
\hline
\end{tabular}
\end{center}
\caption{\label{tb:O2} Possible assignments for $O_2$ of Eq.
(\ref{Opd5}). $N$ denotes the dimension of the lowest dimensional
$\Gamma$ breaking operators.}
\end{table}

We analyzed also $\Gamma$ violating or R-parity violating $O_2$
terms. In this case, we have the lowest order R-parity violating
$O_2$ terms at dimension 8. They are
\begin{align} \label{d=5}
W = & T2_5 T2_5 T6_{\overline{10}} T2_1 \left\{
  \begin{array}{c}
    C^0_7 C^0_4 s^0_5 s^0_8
    +C^-_3 D^+_2 d^+_2 s^-_2 \\
    +C^-_3 D^+_2 d^+_3 s^-_3
    +C^0_7 C^0_6 s^0_5 s^0_3 \\
    +C^0_7 C^0_5 s^0_8 s^0_1
  \end{array}
\right\}.
\end{align}
As commented earlier, the phenomenological constraint on the
coefficient of $O_2$ is rather mild, $<10^{-8}$, and dimension-8
operator can be easily below this bound satisfying the bound on
$O_1$.  Note that all terms in Eqs. (\ref{d=4}) and (\ref{d=5})
include $C_7^0$ or $C_3^-$. So relatively small values of $\langle
C_7^0 \rangle$ and $\langle C_3^-\rangle$ are helpful to fulfill
the constraints.

Therefore, the R-parity violating terms of \cite{Kim:2006hv} can be
made harmless.

Other possible $\Delta B\ne 0$ operators may be present. For
example, we may consider a \PG\ breaking $\langle \overline{\bf
10}^H\rangle\langle \overline{\bf 10}^H\rangle$\bftenb ~\bftenb
~\bffive ~\bffive ~${\bf 5}_{-2}$, $\langle\overline{\bf
10}^H\rangle\langle \overline{\bf 10}^H\rangle$\bftenbh \bftenb~
\bffiveh \bffive ~${\bf 5}_{-2}$, etc.  The former one leads to
dimension-6 operators suppressed by $M_S^4$, and the second one
leads to dimension-5 operators with heavy particle attached. Here,
if any R-parity violating operator occurs, it must involve
sufficient suppression by $M_S$ powers or inclusion of heavy
fields. The baryon number violations with such operators involving
the heavy field \bffiveh\ are safe phenomenologically. It is not
any more dangerous than the standard baryon number violation in
GUTs.

In sum, the parity \PG\ which is a subgroup of an (approximate)
continuous global symmetry is an exact symmetry for operators
involving the MSSM fields only, but is not an exact symmetry of the
full theory. Nevertheless, it is good enough to forbid dangerous
proton decay operators.

There may be $\Delta B=0$ and still \PG\ violating operators. One
well known example is the lepton number violating operators. Since
there is no quadratic superpotential term, \PG\ violation with
$\Delta B=0$ must be looked where superpotential terms have  $D\ge
3$. The phenomenological constraints on $\Delta B=0$ and still
\PG\ violating operators are not so serious \cite{Dreiner}.
Indeed, the lepton number violating operator $lle^+$ in the MSSM
is generated also through $\langle \overline{\bf 10}^H\rangle{\bf
5}{\bf 5}{\bf 1}$ in the flipped SU(5), which is discussed above.
Thus, $lle^+$ is also strongly suppressed.

\section{Conclusion}

  In the $\Z_{12-I}$
model of Ref. \cite{Kim:2006hv}, a discrete parity \PG\ which is a
discrete subgroup of U(1)$_\Gamma$ is found. U(1)$_\Gamma$
contains the U(1)$_X$ symmetry of the flipped SU(5). If the SU(5)
breaking components \bftenbh\ and \bftenh\ are found to carry even
$X$ quantum numbers, then we might have achieved an exact discrete
parity as a subgroup of U(1)$_X$ of flipped SU(5). But in our
model these \bftenbh\ and \bftenh\ carry odd $X$ quantum numbers
and hence we must  resort to an approximate U(1)$_\Gamma$ and the
resulting approximate \PG. Nevertheless, the dangerous dimension-4
and dimension-5 $\Delta B\ne 0$ operators are forbidden up to
sufficiently high dimensions, and the discrete R-parity \PG\ is
harmless. We also noted that this very high order constraint
occurs from the high order 12 of $\Z_{12}$.

\acknowledgments{ This work is supported in part by the KRF Grants
No. R14-2003-012-01001-0, No. R02-2004-000-10149-0 and  No.
KRF-2005-084-C00001. }


\begin{thebibliography}{99}

\def\prp#1#2#3{Phys.\ Rep.\ {\bf #1} (#3) #2}
\def\rmp#1#2#3{Rev. Mod. Phys.\ {\bf #1} (#3) #2}
\def\npb#1#2#3{Nucl.\ Phys.\ {\bf B#1} (#3) #2}
\def\plb#1#2#3{Phys.\ Lett.\ {\bf B#1} (#3) #2}
\def\prd#1#2#3{Phys.\ Rev.\ {\bf D#1} (#3) #2}
\def\prl#1#2#3{Phys.\ Rev.\ Lett.\ {\bf #1} (#3) #2}
\def\jhep#1#2#3{JHEP\ {\bf #1} (#3) #2}
\def\jcap#1#2#3{JCAP\ {\bf #1} (#3) #2}
\def\zp#1#2#3{Z.\ Phys.\ {\bf #1} (#3) #2}
\def\epjc#1#2#3{Euro. Phys. J.\ {\bf C#1} (#3) #2}
\def\ijmp#1#2#3{Int.\ J.\ Mod.\ Phys.\ {\bf #1} (#3) #2}
\def\mpl#1#2#3{Mod.\ Phys.\ Lett.\ {\bf A#1} (#3) #2 }
\def\apj#1#2#3{Astrophys.\ J.\ {\bf #1} (#3) #2}
\def\nat#1#2#3{Nature\ {\bf #1} (#3) #2}
\def\sjnp#1#2#3{Sov.\ J.\ Nucl.\ Phys.\ {\bf #1} (#3) #2}

\bibitem{ibross} L. E. Iba\~nez and G. Ross, \npb{368}{3}{1992}.

\bibitem{Dreiner:2005rd}
  H.~K.~Dreiner, C.~Luhn and M.~Thormeier,
  Phys.\ Rev.\ D {\bf 73} (2006) 075007.

\bibitem{GroHar} D. J. Gross, J. A. Harvey, E. J. Martinec, and R.
Rohm, \prl{54}{502}{1985}.

\bibitem{standlike} L. E. Ibanez, J. E. Kim, H. P. Nilles, and F.
Quevedo, \plb{191}{282}{1987}.

\bibitem{Kim:2006hv}
  J.~E.~Kim and B.~Kyae,
   arXiv:hep-th/0608085 ;
    J.~E.~Kim and B.~Kyae,
  arXiv:hep-th/0608086.

\bibitem{SakYan}N. Sakai and T. Yanagida, \npb{197}{533}{1982};
S. Weinberg, \prd{26}{287}{1982}.

\bibitem{anomalousU(1)}  H.~K.~Dreiner, H.~Murayama and M.~Thormeier,
  Nucl.\ Phys.\ B {\bf 729} (2005) 278.  See also
  H.~K.~Dreiner, C.~Luhn, H.~Murayama and M.~Thormeier,
  arXiv:hep-ph/0610026.


\bibitem{discreteg} L. Krauss and F. Wilczek,
 \prl{62}{1221}{1989};
 T. Banks, \npb{323}{90}{1989};
   S.~P.~Martin,
  Phys.\ Rev.\ D {\bf 46} (1992) 2769.

\bibitem{tHooft} G. 't Hooft, \npb{35}{167}{1971}.

\bibitem{ChKimKim} K.-S. Choi, I.-W. Kim and J. E. Kim,
hep-ph/0612107.

\bibitem{Dreiner} B. C. Allanach, A. Dedes and H. K. Dreiner,
\prd{60}{075014}{1999}.

\end{thebibliography}
\end{document}